# Improved performance of a near-field thermophotovoltaic system by a back gapped reflector

Dudong Feng, Shannon K. Yee, and Zhuomin M. Zhang*
The George W. Woodruff School of Mechanical Engineering
Georgia Institute of Technology, Atlanta, GA 30332, USA

*Corresponding Author: zhuomin.zhang@me.gatech.edu

**ABSTRACT**

Various spectral control techniques can be applied to improve the performance of a thermophotovoltaic (TPV) system. For example, a back surface reflector (BSR) can improve the performance of TPV systems. A conventional metal BSR structure enhances the photogeneration rate by increasing the absorption probability of photons via back surface reflection, affording a second chance for absorption. However, surface passivation and external luminescence effects introduced by BSR structures have been previously ignored, which potentially decreases the performance of TPV systems. Recently, a back gapped reflector (BGR) structure was proposed to greatly improve the performance of far-field TPV systems by reducing reflection loss at the semiconductor-metal interface. In the present work, the performance improvement on a thin-film, near-field InAs TPV system with a BGR is investigated, comparing its performance to that with a conventional metal BSR. Surface passivation conditions are also investigated to further improve the performance of TPV systems with back reflectors. The output power and efficiency are calculated using an iterative model combining fluctuational electrodynamics and the full drift-diffusion model. For the well-passivated condition, when the BSR is replaced by the BGR, the calculated conversion efficiency was improved from 16.4% to 21% and the output power was increased by 10% for the near-field regime. Finally, the reflection loss and external luminescence loss are analyzed to explain the performance improvement.

## 1. Introduction

Thermophotovoltaics (TPVs) convert thermal radiation into electricity from terrestrial heat sources. A TPV system consists of three main components: an emitter that radiates photons by consuming thermal energy from a heat source, a TPV cell that converts incident photons into electricity, and a distance-tunable vacuum gap between the emitter and cell. With the benefits of compact size and solid-state operation, TPV systems could achieve widespread applications enabling local thermal energy storage [1, 2], waste heat recovery [3, 4], aerospace power generation [5, 6], nuclear power conversion [7, 8], and direct solar energy conversion [9-16]. Among various applications, solar thermophotovoltaics (STPVs) is a promising electricity generating technology that possesses a high theoretical system efficiency of 54% without using concentration or a selective absorber [17], and exceeds the Shockley-Queisser limit of 33.7%.

Datas and Algora [16] developed a practical STPV device using HfO$_2$ coated tungsten as the absorber and emitter, which exhibited 0.8% system efficiency. Lenert *et al*. [10] experimentally demonstrated a multiwalled carbon nanotube STPV device with 3.2% efficiency. Bierman *et al*. [15] further improved the solar-to-electrical efficiency reaching 6.8% with a one-dimensional photonic crystal selective emitter and a tandem plasma-interference optical filter. The primary difference between the practical efficiency and theoretical efficiency of STPV devices suggests that even greater performance can be achieved.

Current efforts have been directed toward optimizing the photon emission and absorption processes to match the bandgap energy of cell materials, which is the primary approach to realizing high efficiency TPV systems [17-19]. Several research groups have achieved wavelength selectivity and direction-insensitivity on TPV emitters using periodic nanostructures including gratings [20-23], nanowires [24], and photonic crystals [25], which also exhibit enhanced emittance by the excitations of surface polaritons and cavity resonance. Considering the working temperature of TPV systems (*e.g*., 600-1300 K), narrow bandgap semiconductor materials are preferred to produce larger output power density than regular solar cell materials [4, 14, 26]. Additionally, near-field radiation has been proposed as one of the promising mechanisms to greatly enhance the performance of TPV systems by Whale and Carvalho [27]. When the tunable vacuum gap distance is reduced to the microscale or even nanoscale, additional coupled evanescent photons tunnel through the gap and dominate the radiation exchange between the emitter and cell [28]. Significant experimental progress has been made in recent years [28-32]. To further optimize the efficiency and boost the output power density of near-field TPV systems, the similar spectral control techniques mentioned above can also be applied to various systems [33-36].

A back reflector is a common way to improve the performance of a TPV system by increasing the interband absorption while recycling photons below bandgap energy [37, 38]. Two different BSRs were investigated for a far-field Ge TPV cell by Fernandez *et al*. [39]. The proposed dielectric BSR can feature high infrared reflection and excellent surface passivation of the backside of Ge TPV cells; however, the effect of the parasitic absorption of BSRs on the TPV performance has not been discussed. A parametric study investigating the effect of BSR and surface recombination on a near-field InGaSb TPV systems was performed by Bright *et al.* [38], who investigated the effect of a gold BSR and surface recombination separately on a near-field InGaSb TPV system. This parametric study used a combined model that neglected the near-field effects on the dark current and internal luminescence, which might be important for certain TPV systems [40, 41]. Recently, Fan *et al.* [42] proposed a thin-film TPV cell design with a BGR, which can greatly enhance the performance of a far-field InGaAs TPV system by reducing the parasitic absorption in the back reflector. Similar structures have not been applied to near-field thin-film TPV systems, which may see even greater improvements in system performance.

In this paper, a thin-film InAs TPV system with a BGR is investigated using a newly proposed photon-charge coupled iterative model [41]. By comparing to a conventional metal BSR with different surface passivation conditions, the performance enhancement potential of this BGR is investigated for both the far- and near-field regimes. The radiation exchange carried by propagating waves and evanescent waves are separately discussed for different vacuum spacings.

The thickness of the BGR is varied to optimize the efficiency of the InAs TPV system. The efficiency reduction, due to reflection at the metal-semiconductor interface, and output power loss, due to external luminescence and surface recombination, are also analyzed. Note, the loss mechanisms also depend on the gap thickness of the BGR, which can potentially guide the design and optimization of higher efficiency TPV systems with larger output power.

## 2. Methodology

A thin-film InAs TPV system with a BGR is shown in Fig. 1a, which is compared with a TPV system with a conventional metal BSR shown in Fig. 1b. A tunable vacuum spacing ($d$) separates the tungsten emitter and InAs TPV cell. The tungsten emitter is modeled as an infinitely thick bulk with temperature at $T_e = 900$ K. The $p$-doped and $n$-doped InAs layers with thicknesses of $d_p = d_n = 200$ nm, are respectively subdivided by a 50-layer nonuniform mesh to consider the local effect on photogeneration and photon chemical potential [41]. Au is used for both BGR and BSR, and the thickness $d_{Au}$ is chosen to be 100 nm to ensure all incident photons are reflected or absorbed. The front metal grid and the back metal grid are ignored in the model due to the low shading area in the radiation exchange process. The back metal grid in the BGR structure is used to support the vacuum gap between the cell and the back reflector. The temperatures of the InAs cell (including BSRs or BGRs) are assumed to be room temperature, $T_c = 300$ K. Both systems require a vacuum environment to minimize the convection between the emitter and cell.

The charge transport equations for each layer are prescribed by full drift-diffusion model expressed as [41]:

$$\nabla \cdot (\varepsilon_r \nabla \varphi) = (q/\varepsilon_0)(N_A - N_D + n - p) \tag{1}$$

$$\frac{\partial n}{\partial t} = \frac{1}{q} \nabla \cdot J_e + g - r_{Auger} - r_{SRH} \tag{2}$$

$$\frac{\partial p}{\partial t} = -\frac{1}{q} \nabla \cdot J_h + g - r_{Auger} - r_{SRH} \tag{3}$$

where $\varepsilon_r$ is the dielectric constant of the cell material, $\varepsilon_0$ is the vacuum permittivity, $q$ is the elementary charge, and $N_A$ and $N_D$ are the acceptor and donor concentrations, respectively. Note that $n$ and $p$ are the electron and hole concentrations, which are functions of time and space.

The net photogeneration rate per unit volume is uniformly distributed inside layer $j$, which is solved by fluctuational electrodynamics with 100% internal quantum efficiency:

$$g = \sum_{m \neq j} \int_{\omega_g}^{\infty} \left[ \Psi(\omega, T_m, \mu_m) - \Psi(\omega, T_j, \mu_j) \right] \Upsilon_{mj}(\omega) \frac{d\omega}{\Delta z_j} \tag{4}$$

where $\omega$ is the angular frequency, $\omega_g = E_g/\hbar$ is the frequency that corresponds to the bandgap energy ($E_g$) with $\hbar$ being the reduced Planck constant, and $T$ and $\mu$ are the absolute temperature and photon chemical potential of the corresponding layer, respectively. $\Delta z_j$ is the thickness of layer $j$. The function $\Psi$ is the modified Bose-Einstein distribution [41]. The function $\Upsilon_{mj}$ is the fraction of photons emitted at a given frequency from layer $m$ that is absorbed by layer $j$ and vice versa. The detailed expression can be found in Ref. [41, 43]. To be mentioned, the external luminescence from the tiny back region (less than the parallel cutoff wavelength) of $n$-InAs layer to BSRs is ignored, which means only the local optical response is considered [44, 45].

The Auger recombination and the Shockley-Read-Hall (SRH) recombination rate are calculated respectively as follows [41]:

$$r_{\text{Auger}} = (C_e n + C_h p)(np - n_i^2) \tag{5}$$

and

$$r_{\text{SRH}} = \frac{np - n_i^2}{\tau_p(n + n_{t,b}) + \tau_n(p + p_{t,b})} \tag{6}$$

where $C_e$ and $C_h$ are the Auger recombination coefficients for electrons and holes, $n_i$ is the intrinsic carrier concentration, $\tau_n$ and $\tau_p$ are the bulk lifetimes for electrons and holes, respectively. $n_{t,b}$ and $p_{t,b}$ are the electron and hole trap concentrations that are set to be the same as the $n_i$ in the modeling.

The charge current densities, $J_e$ and $J_h$, are modeled in terms of drift and diffusion forces as follows:

$$J_e = -q\upsilon_e n \nabla \varphi + q D_e \nabla n \tag{7}$$

$$J_h = -q\upsilon_h p \nabla \varphi + q D_h \nabla p \tag{8}$$

where $\upsilon_e$ and $\upsilon_h$ are the mobility of electrons and holes, respectively, and $D_e$ and $D_h$ are the diffusion coefficients, which are related to mobility according to Einstein's relation $D = \upsilon k_B T / q$ for each type of carrier. The electron and hole mobilities are functions of temperature and doping concentration, which is calculated by a Caughey-Thomas-like model [41, 46]. Ohmic contact is assumed and governing equations (1), (2), and (3) are solved by applying the following boundary conditions:

$$J_{(e,h)}(z_s) = q S_{(e,h),(n,p)}[n(z_s) - n_0(z_s)] \tag{9}$$

where $S_{e,p}$, $S_{h,p}$, $S_{e,n}$, and $S_{h,n}$ are the surface recombination velocities for electrons and holes in the $p$ and $n$ regions, respectively, and $n_0$ and $p_0$ are the carrier concentrations at equilibrium. $z_s$ represents the location at the top surface of $p$-InAs and bottom surface of $n$-InAs.

Equation (4) is a function of the spatial profile of photon chemical potential, which is the solution of Eqs. (1), (2), and (3). Therefore, an iterative method is required to solve this coupled transport problem. The general procedure of this iterative solver can be found in Ref. [41].

The net rate of absorbed energy of layer $j$ from the tungsten emitter can be calculated by

$$Q_j = \sum_m^{\text{emitter}} \int_0^\infty \hbar\omega \left[ \Psi(\omega, T_m, \mu_m) - \Psi(\omega, T_j, \mu_j) \right] \Upsilon_{mj}(\omega) d\omega \qquad (10)$$

The total absorbed energy for a practical TPV system should include the absorption by the back reflector, since extra work is required to remove this parasitic heating. Therefore, the conversion efficiency is calculated by

$$\eta(V) = \frac{J(V)V}{Q_{\text{cell}}}, \text{ where } Q_{\text{cell}} = \sum_j^{\text{cell}} Q_j \qquad (11)$$

Here, the summation is over the InAs cell. The absorbed energy by the Au layer is taken into account in calculating the efficiencies.

## 3. Results and Discussion

The dielectric function of tungsten at $T_e$ is assumed to be identical to that at room temperature [47]. The dielectric function of InAs is taken from Ref. [48] without considering the free-carrier contributions. A Drude model is used to describe the dielectric function of Au [28], $\varepsilon(\omega) = \varepsilon_\infty - \omega_p^2/(\omega^2 + i\gamma\omega)$, where $\varepsilon_\infty = 1$, $\omega_p = 1.37 \times 10^{16}$ rad/s, and $\gamma = 5.31 \times 10^{13}$ rad/s [38]. The doping concentrations of acceptor and donor are set as $N_A = 8 \times 10^{17}$ cm$^{-3}$ and $N_D = 2 \times 10^{16}$ cm$^{-3}$, respectively [41]. The intrinsic concentration is $n_i = 6.06 \times 10^{14}$ cm$^{-3}$ for InAs at room temperature. For Auger and SRH recombination, the following parameters are used: $C_e = C_h = 2.26 \times 10^{-27}$ cm$^6$ s$^{-1}$ and $\tau_e = \tau_h = 100$ ns [49]. The values for mobilities in $p$- and $n$- regions are as follows: $\upsilon_{e,p} = 18300$ cm$^{-2}$ V$^{-1}$ s$^{-1}$, $\upsilon_{h,p} = 166$ cm$^{-2}$ V$^{-1}$ s$^{-1}$, $\upsilon_{e,n} = 26800$ cm$^{-2}$ V$^{-1}$ s$^{-1}$, and $\upsilon_{h,n} = 370$ cm$^{-2}$ V$^{-1}$ s$^{-1}$ [41].

In the following, the vacuum spacing is chosen as $d = 1$ mm or $d = 10$ nm to investigate the effect of this BGR on the TPV system under either the far- or near-field regime, respectively. The gap thickness, $h$, is parametrically swept between 10 nm to 10 μm. The efficiency and maximum output power are calculated as a function of the thickness of the BGR for both working regimes. Surface passivation conditions can potentially affect the performance of TPV systems [38, 39, 50, 51]. A well-passivated surface usually has large surface recombination velocity for majority carriers and small surface recombination velocity for minority carriers [52]. The surface recombination velocity for majority carriers is set to be an extremely large number, e.g., $10^9$ cm/s. The surface recombination velocity for minority carriers can vary from $10^2$ cm/s to $10^4$ cm/s or larger depending on surface passivation conditions [49]. In this paper, $10^4$ cm/s and $10^2$ cm/s are used to represent a non-passivated surface and a passivated surface, respectively. By comparing the BGR with the BSR under the same passivation condition, the effect of external luminescence loss can be explicitly distinguished from that of surface passivation. The front surfaces of four calculations are all assumed to be well-passivated. The surface recombination velocities for each TPV system are listed in Table 1.

Table 1. Surface recombination velocities for each TPV cell

| TPV systems | $S_{e,p}$ (cm/s) | $S_{h,p}$ (cm/s) | $S_{e,n}$ (cm/s) | $S_{h,n}$ (cm/s) |
|---|---|---|---|---|
| Passivated BSR/BGR | $10^2$ | $10^9$ | $10^9$ | $10^2$ |
| Non-passivated BSR/BGR | $10^2$ | $10^9$ | $10^9$ | $10^4$ |

*3.1. Output electric power and photogeneration rate*

The maximum output power and photogeneration rate of four TPV systems at two different vacuum spacings are shown in Fig. 2 and Fig. 3. As is clearly shown in Fig. 2a, the maximum output power of the TPV system with a BGR slightly increases to a saturated value as the gap thickness increases. The enhancement effect on the output power of the TPV system with a BGR is demonstrated by comparison with a conventional metal BSR. The maximum output power of the BGR scenario is 15%-40% higher, compared to the BSR scenario for the passivated condition. Although the maximum output power of the non-passivated condition is reduced by almost one third of the passivated condition, the maximum output power of the BGR TPV is 5%-15% larger than that of the BSR TPV for the passivated condition. Enhanced total net photogeneration rate and a well-passivated surface can greatly improve the maximum output power of a TPV system. Shown in Fig. 2b, the total net photogeneration rate of the BGR exhibits a similar trend as Fig. 2a. As the total net photogeneration rate increases with the gap thickness, more free electrons and holes are generated in the TPV cell, which produces a larger output power than that of BSR. The total net photogeneration rate of the BGR is 10%-25% greater than the passivated BSR and 3%-15% larger than the non-passivated BSR. The enhancement effect on the total net generation rate of the passivated BGR is lower than that on the maximum output power, which demonstrates that the enhancement on the total net photogeneration rate is not reflected linearly on maximum output power for the passivated scenario. The total net photogeneration rate can also be expressed by subtracting the external luminescence from the total net photogeneration rate between the emitter and the *pn* junction. A good surface passivation would result in a larger spatial profile of photon chemical potential, which could possibly enhance the external luminescence. Therefore, the total net photogeneration rate of the non-passivated BSR is larger than that of the passivated BSR. The spatial profile of the net photogeneration rate shown in Fig. 2c indicates a coherent absorption pattern of propagating waves inside the semiconductor. The spatial profile of the net photogeneration rate of the passivated BSR results from a combined effect of strong external luminescence and spatial distribution of photon chemical potential. For the non-passivated scenario, the BSR has a similar profile with the 10 nm BGR because the photon tunneling possibilities from the *pn* junction to metal for both scenarios are similar. As getting closer to the backside surface of *pn* junction, the growing divergence of the net photogeneration rate between the BSR and BGR indicates that the external luminescence of the BSR is much larger than that of the 10 nm BGR. As the gap thickness increases, the peak of coherent absorption pattern shifts and the total photogeneration rate increases around 10% for the non-passivated BGR. Nevertheless, the BGR structure with good passivated surfaces can definitely enhance the maximum output power at the far-field regime for different gap thicknesses.

For the near-field condition shown in Fig. 3a, the maximum output power of the BGR decreases as the gap thickness increases for both passivated and non-passivated conditions. Whether the BGR structure can provide enhancement on the maximum output power compared to the BSR structure depends on the gap thickness. The largest enhancement from the BGR structure is when the gap thickness equals 10 nm. The maximum output power generated by the BGR has a 10% enhancement compared to that of the passivated BSR. In addition, the maximum improvement of the BGR structure is reduced to 2.5% for the non-passivated condition. The surface passivation effect can still be clearly seen by the difference of maximum output power between the passivated and non-passivated scenario. The trend of the maximum output power of the BGR matches well with the total net photogeneration rate, shown in Fig. 3b. The corresponding gap thickness at the intersections of the BGRs and the BSRs in Fig. 3a are the same with those in Fig. 3b. Unlike the far-field condition, the reduction of the total net photogeneration rate as the gap thickness increases is not only due to the propagating photons but also the frustrated photons for near-field condition. As shown in Fig. 3c, the spatial absorption peak exists in the front surface of the TPV cell because large frequency photons can only penetrate a skin depth of lossy materials [28]. Since most energy is carried by propagating and frustrated photons, the photogeneration rate reduction happens at a location close to the back surface, which can be clearly seen in Fig. 3c. The main reason for this photogeneration reduction is because the coherent absorption pattern in the middle of the semiconductor becomes destructive for certain wavelengths. For the near-field condition, a BGR with a small gap thickness is preferred for larger output power.

*3.2. Efficiency and cell absorbed energy*

The maximum conversion efficiency improvement brought by the BGR structure is clearly shown in Fig. 4. The BGR exhibits improvement to the maximum conversion efficiency of the TPV systems. As seen in Fig. 4a, the conversion efficiency using a BGR is 1.05-1.34 times higher than that using a BSR for non-passivated condition. The passivated BGR maximum efficiency ranges between 5.7%-7.4%, which is a 1.18-1.62 times higher than the passivated BSR when the gap thickness increases. This enhancement in efficiency is higher than that of the output power (15%-40%) indicating that the efficiency improvement is a coupled effect of the output power enhancement and parasitic absorption reduction. As clearly shown in Fig. 4b, the cell absorbed energy is inversely correlated with the maximum efficiency as the gap thickness increases. The reduction effect on cell absorbed energy can be clearly seen by the difference between the BGR and BSR.

For the near-field condition shown in Fig. 5a, the passivated BGR has an 18%-21% maximum conversion efficiency when the gap thickness increases, which is higher than the 16.4% for the passivated metal BSR. For the non-passivated scenario, the BGR also possesses a 1.08-1.19 times improvement when compared with the BSR. As manifest in Fig. 5b, the maximum efficiency increases by increasing the gap thickness. Since the maximum output power is not a strong function of the gap thickness, the cell absorbed energy should be inversely proportional to gap thickness to match the efficiency enhancement. The cell absorbed energy reduces as the gap thickness increases in Fig. 6b. Compared to the propagating modes, the cell absorbed energy carried by the frustrated modes (the difference between the purple and blue curves) is largely reduced in the near-field scenario. As the vacuum spacing decreases from Fig. 5b ($d$ = 1 mm) to Fig. 6b ($d$ = 10 nm), near-field radiation dominates the radiation exchange between the emitter and cell.

*3.3. Reflection loss and external luminescence loss*

The generated power can be enhanced by adding a conventional metal BSR to a TPV cell [34, 38, 39]. An ideal back reflector reflects all incident photons. However, a perfect mirror for all wavelengths of photons does not exist. Therefore, the parasitic absorption, known as the reflection loss of the back reflector, is difficult to eliminate, requiring extra cooling power to remove the parasitic absorbed energy. Therefore, it is important to minimize the reflection losses. When light transmits through the semiconductor-metal interface of a conventional metal BSR structure, the reflection is lossy because most of the electromagnetic waves can easily transmit from an optically rare medium to an optically dense medium. However, by adding a gap between the semiconductor and back reflector, the photons reflect perfectly at the semiconductor-air interface through total internal reflection. However, if the gap thickness is comparable to or less than the wavelength of transmitted photons, the photon tunneling effect would also be present between the semiconductor and metal. Therefore, the gap thickness is required to be larger than the characteristic wavelength to avoid photon tunneling effects and to minimize the energy absorption of the back reflector. In Fig. 6, it is shown that the parasitic absorption in the BGR decreases greatly as the gap thickness increases for both far- and near-field regimes. For the near-field condition shown in Fig. 6b, the frustrated modes absorbed by the metal in the BGR structure (the difference between the purple and blue curves) are perfectly reduced to zero by increasing the gap thickness. The difference of $Q_{metal}$ between a passivated BSR and non-passivated BSR is due to the external luminescence induced by the different photon chemical potential, which will be discussed in the next paragraph. The spectral absorbed energy of the back reflectors ($q_{metal}$) of the passivated BSR, non-passivated BSR, and selected BGRs ($h$ = 10 nm, 1 μm and 10 μm) are shown in Fig. 7 for the TPV systems with a vacuum spacing of $d$ = 10 nm. As the gap thickness increases, the absorbed energy of the back reflectors reduces for every angular frequency. Also, the reduction of spectral absorbed energy of back reflectors decreases as the angular frequency increases because a smaller gap thickness is required for photons with larger energy to tunnel through. By replacing the conventional metal BSR with a BGR, the parasitic absorption inside the back reflector is greatly reduced and the conversion efficiency of TPV systems can be potentially improved.

In a working TPV cell, the more electron and hole pairs that are generated, the more likely the electron and hole pairs will recombine and generate external luminescence, similar to a biased LED [40, 41, 53]. The luminescent intensity from a working TPV cell is exponentially proportional to the photon chemical potential, which is usually assumed to be equal to the product of the working voltage and elementary charge. The energy of the luminescent photons from the cell cannot be recycled when they are absorbed by any object other than the emitter and cell; this is called external luminescence loss. As one of the major loss mechanisms that deteriorates the performance of TPV systems working at moderate temperatures, the external luminescence loss can also be minimized by the BGR structure. As shown in Fig. 8a, at the maximum output power voltage ($V_{max}$), the TPV cell with the passivated BSR structure at $V_{max}$ = 0.054 V would emit ~10% photons of the total photogeneration rate. Due to the dominant effects of surface recombination in the non-passivated BSR, $V_{max}$ is only 0.022 V, which is much less than that of the passivated BSR. Nevertheless, there is still 2% of external luminescence loss that exists in the non-passivated BSR. However, the external luminescence loss can be reduced to 1% of the total net photogeneration rate by introducing a 10 nm gap between the semiconductor and the back reflector. As the gap thickness increases, the external luminescence loss ratio can be reduced to less than 0.01%. Similar trends are exhibited in Fig 8b, and a 5.6% external luminescence loss can be reduced to effectively zero by the BGR structure for both the passivated and non-passivated condition. Although $V_{max}$ increases when the vacuum spacing reduces, the BGR structure can still reduce the external luminescence losses to negligible levels. As the emitter temperature increases, the TPV systems will be working at a higher voltage, which means a larger external luminescence would occur. Therefore, minimizing the external luminescence loss is of great significance in creating a high-performance TPV system.

4. **Conclusions**

This work theoretically demonstrates the performance improvement of a thin-film InAs TPV system by replacing the conventional metal BSR with a BGR. A coupled photon-charge transport model that accounts for surface recombination and external luminescence is used to model these two TPV systems for both far- and near-field regimes. The performance improvement due to this BGR structure is addressed when compared with the passivated BSR. The results show that the output power and system efficiency are significantly improved for far-field TPV systems with a BGR; this is consistent with the previous study using an InGaAs cell. Moreover, the near-field InAs TPV systems with a BGR also exhibit a 10% improvement of the output power and an efficiency boost of 4.5% (absolute value) over the BSR under well-passivation condition. The performance enhancement brought by the well-passivated back surface is huge for both the BSR and BGR. The reflection loss is significantly reduced by adding a gap between the semiconductor and back reflector for both far- and near-field TPV systems. The effect of external luminescence on the total net photogeneration rate can be neglected in the TPV system with a BGR. The parametric sweeps on the gap thickness of the BGR provide a clear direction for the design and optimization for the back reflectors for TPV systems. The same method may be applied to studying the effect of a BGR structure on different TPV systems.

**Acknowledgement**

This work was supported by the U.S. Department of Energy, Office of Science, Basic Energy Sciences (DE-SC0018369).

**Data availability**

The data that support the findings of this study are available from the corresponding authors upon reasonable request.

[15]     D. M. Bierman, A. Lenert, W. R. Chan, B. Bhatia, I. Celanović, M. Soljačić, and E. N. Wang, Enhanced photovoltaic energy conversion using thermally based spectral shaping, Nat. Energy 1 (2016) 16068.

[16]     A. Datas and C. Algora, Development and experimental evaluation of a complete solar thermophotovoltaic system, Prog. Photovoltaics Res. Appl. 21 (2013) 1025-1039.

[17]     N.-P. Harder and P. Wurfel, Theoretical limits of thermophotovoltaic solar energy conversion, Semicond. Sci. Tech. 18 (2003) S151-S157.

[18]     S. Basu, Y.-B. Chen, and Z. M. Zhang, Microscale radiation in thermophotovoltaic devices—A review, Int. J. Energy Res. 31 (2007) 689-716.

[19]     T. Bauer, Thermophotovoltaics: Basic Principles and Critical Aspects of System Design, Springer, Berlin, 2011.

[20]     Y. B. Chen and Z. M. Zhang, Design of tungsten complex gratings for thermophotovoltaic radiators, Opt. Commun. 269 (2007) 411-417.

[21]     B. Zhao, L. Wang, Y. Shuai, and Z. M. Zhang, Thermophotovoltaic emitters based on a two-dimensional grating/thin-film nanostructure, Int. J. Heat Mass Transfer 67 (2013) 637-645.

[22]     H. Sai, Y. Kanamori, and H. Yugami, Tuning of the thermal radiation spectrum in the near-infrared region by metallic surface microstructures, J. Micromech. Microeng. 15 (2005) S243-S249.

[23]     L. P. Wang and Z. M. Zhang, Wavelength-selective and diffuse emitter enhanced by magnetic polaritons for thermophotovoltaics, Appl. Phys. Lett. 100 (2012) 063902.

[24]     V. Tomer, R. Teye-Mensah, J. C. Tokash, N. Stojilovic, W. Kataphinan, E. A. Evans, G. G. Chase, R. D. Ramsier, D. J. Smith, and D. H. Reneker, Selective emitters for thermophotovoltaics: erbia-modified electrospun titania nanofibers, Sol. Energy Mater. Sol. Cells 85 (2005) 477-488.

[25]     A. Narayanaswamy and G. Chen, Thermal emission control with one-dimensional metallodielectric photonic crystals, Phys. Rev. B 70 (2004) 125101.

[26]     D. Cakiroglu, J. P. Perez, A. Evirgen, C. Lucchesi, P. O. Chapuis, T. Taliercio, E. Tournié, and R. Vaillon, Indium antimonide photovoltaic cells for near-field thermophotovoltaics, Sol. Energy Mater. Sol. Cells 203 (2019) 110190.

[27]     M. D. Whale and E. G. Cravalho, Modeling and performance of microscale thermophotovoltaic energy conversion devices, IEEE Trans. Energy Conv. 17 (2002) 130-142.

[28]     Z. M. Zhang, Nano/Microscale Heat Transfer, 2nd ed., Springer Nature, Cham, 2020.

[29]     M. Ghashami, A. Jarzembski, M. Lim, B. J. Lee, and K. Park, Experimental exploration of near-field radiative heat transfer, Ann. Rev. Heat Transfer 23 (2020) 13-56.
11

**Figure captions**

Fig. 1.  Schematic of an InAs TPV system with (a) a back gapped reflector (BGR) and (b) a conventional BSR. Here, $d$ is the vacuum spacing between the emitter and the cell, and $h$ is the back gap thickness.

Fig. 2.  (a) Maximum output power, (b) total net photogeneration rate, and (c) spatial profile of the photogeneration rate as a function of the gap thickness for TPV systems with different back reflectors and passivation conditions when $d = 1$ mm. The total net photogeneration rate and the spatial profile of the photogeneration rate are both calculated at the maximum output power condition for each TPV system. The passivated BGR is similar to the non-passivated BGR, therefore, only the non-passivated BGR with three gap thickness are shown in Fig. 2b and 2c.

Fig. 3.  (a) Maximum output power, (b) total net photogeneration rate, and (c) spatial profile of the photogeneration rate as a function of the gap thickness for TPV systems with different back reflectors and passivation conditions when $d = 10$ nm. The total net photogeneration rate and the spatial profile of the photogeneration rate are both calculated at the maximum output power condition for each TPV system. The passivated BGR is similar to the non-passivated BGR, therefore, only the non-passivated BGR with three gap thickness are shown in 3b and 3c.

Fig. 4.  (a) Maximum conversion efficiency and (b) cell absorbed energy as a function of the gap thickness for TPV systems with different back reflectors and passivation conditions when $d = 1$ mm. The surface passivation has little effect on the cell absorbed energy, therefore, the BGR and the BSR include both the passivated and non-passivated conditions in Fig. 4b.

Fig. 5.  (a) Maximum conversion efficiency and (b) cell absorbed energy as a function of the gap thickness for TPV systems with different back reflectors and passivation conditions when $d = 10$ nm. The surface passivation has little effect on the cell absorbed energy, therefore, the BGR and the BSR include both the passivated and non-passivated conditions in Fig. 5b.

Fig. 6.  Metal absorbed energy ($Q_{metal}$) as a function of the gap thickness for TPV systems with different back reflectors when (a) $d = 1$ mm and (b) $d = 10$ nm.

Fig. 7.  Spectral absorbed energy of the metal ($q_{metal}$) of different near-field TPV systems with different back reflectors when $d = 10$ nm.

Fig. 8.  External luminescence loss ratio as a function of the gap thickness for TPV systems with different back reflectors when (a) $d = 1$ mm and (b) $d = 10$ nm at the maximum output power voltage. The external luminescence over the total net photogeneration rate is called external luminescence loss ratio.

(a) 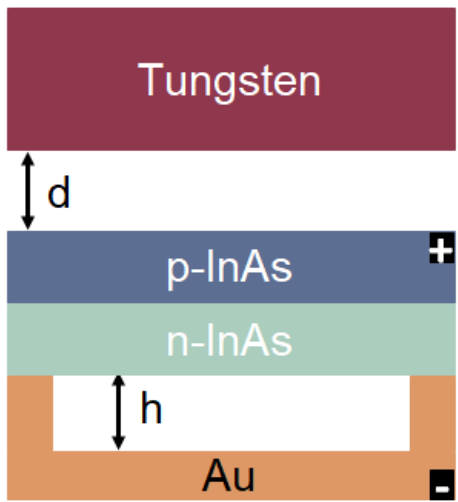
(b) 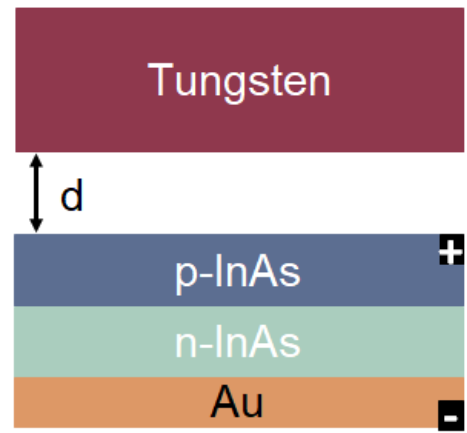

Fig. 1, Feng et al.

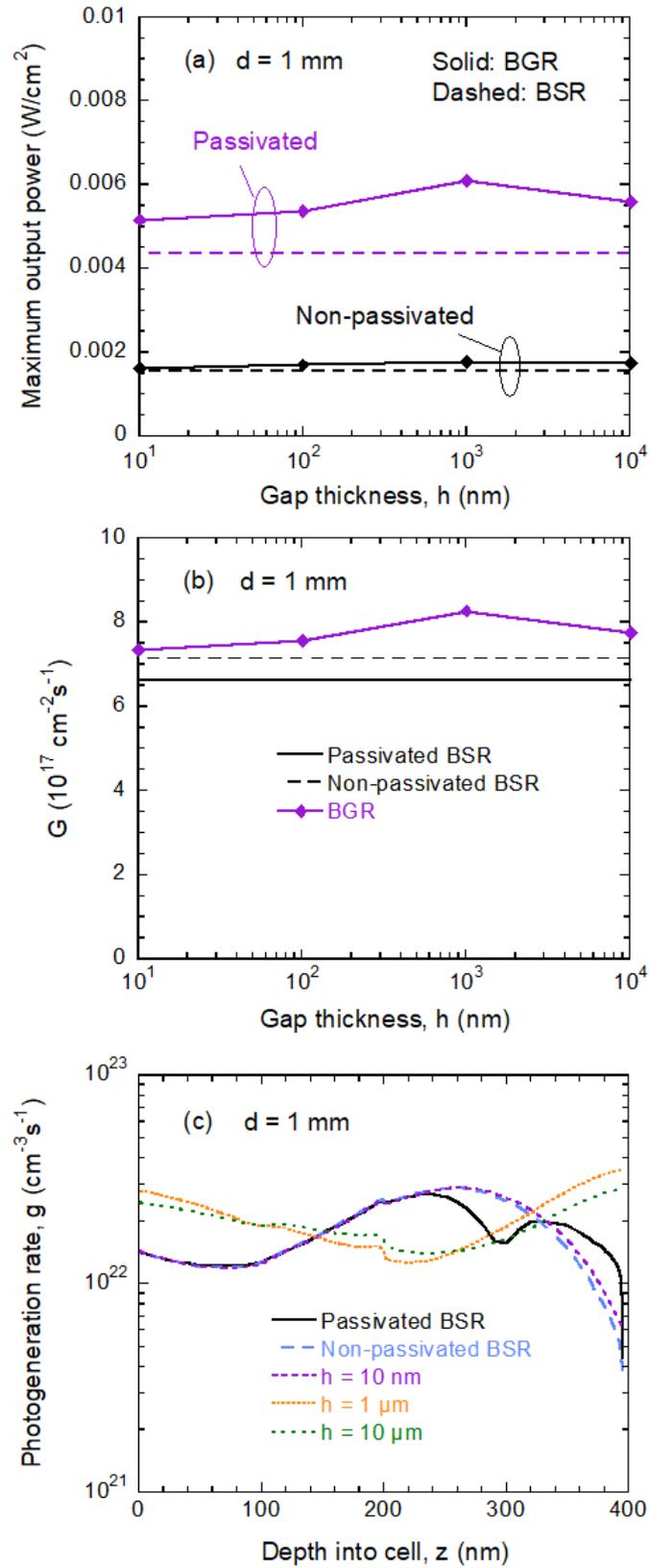

Fig. 2, Feng et al.

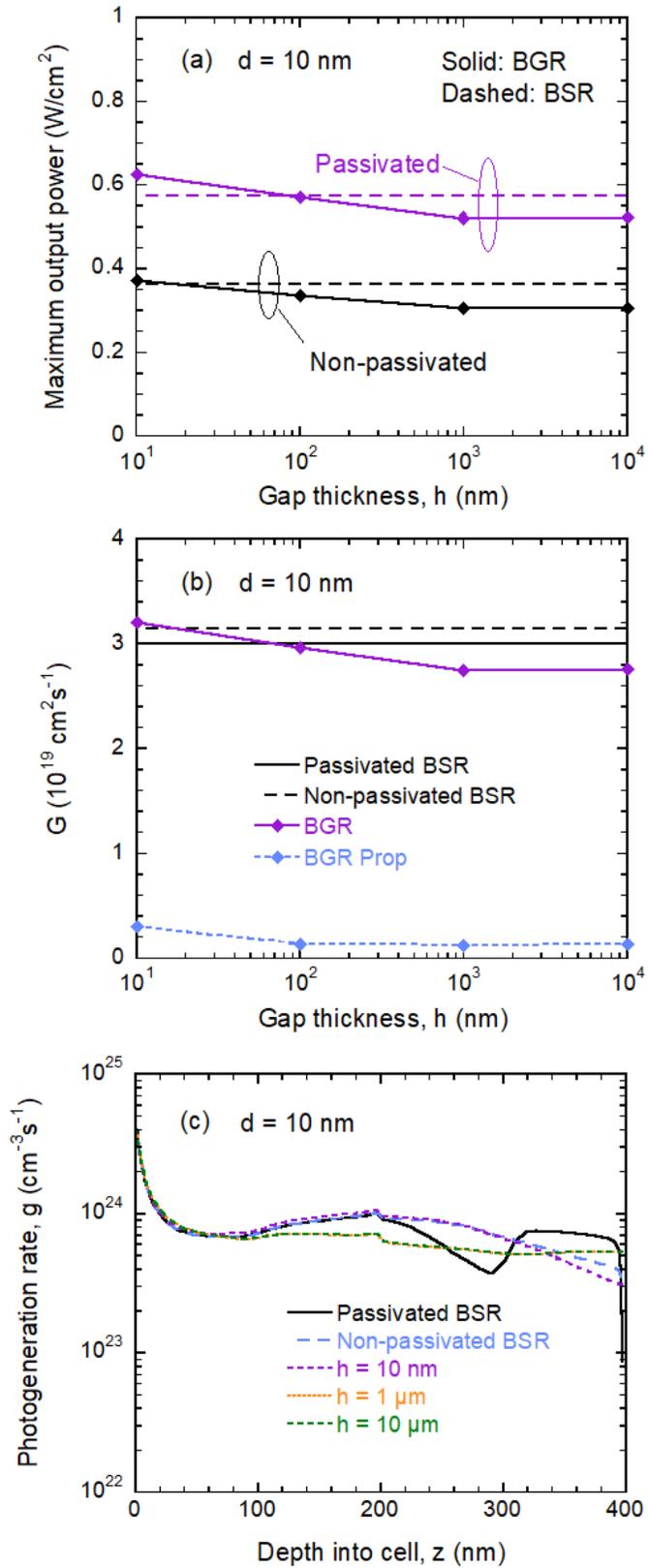

Fig. 3, Feng et al.

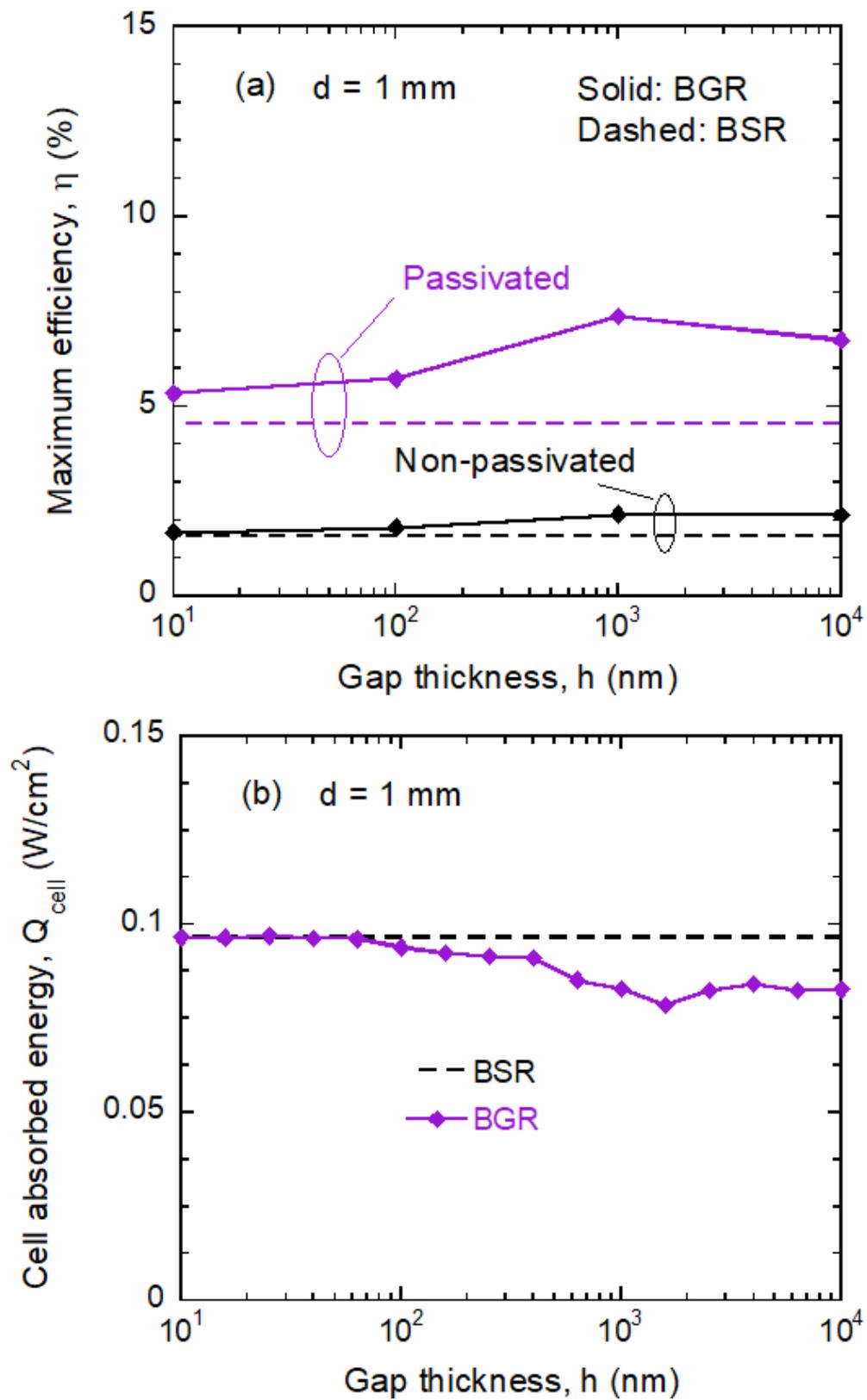

Fig. 4, Feng et al.

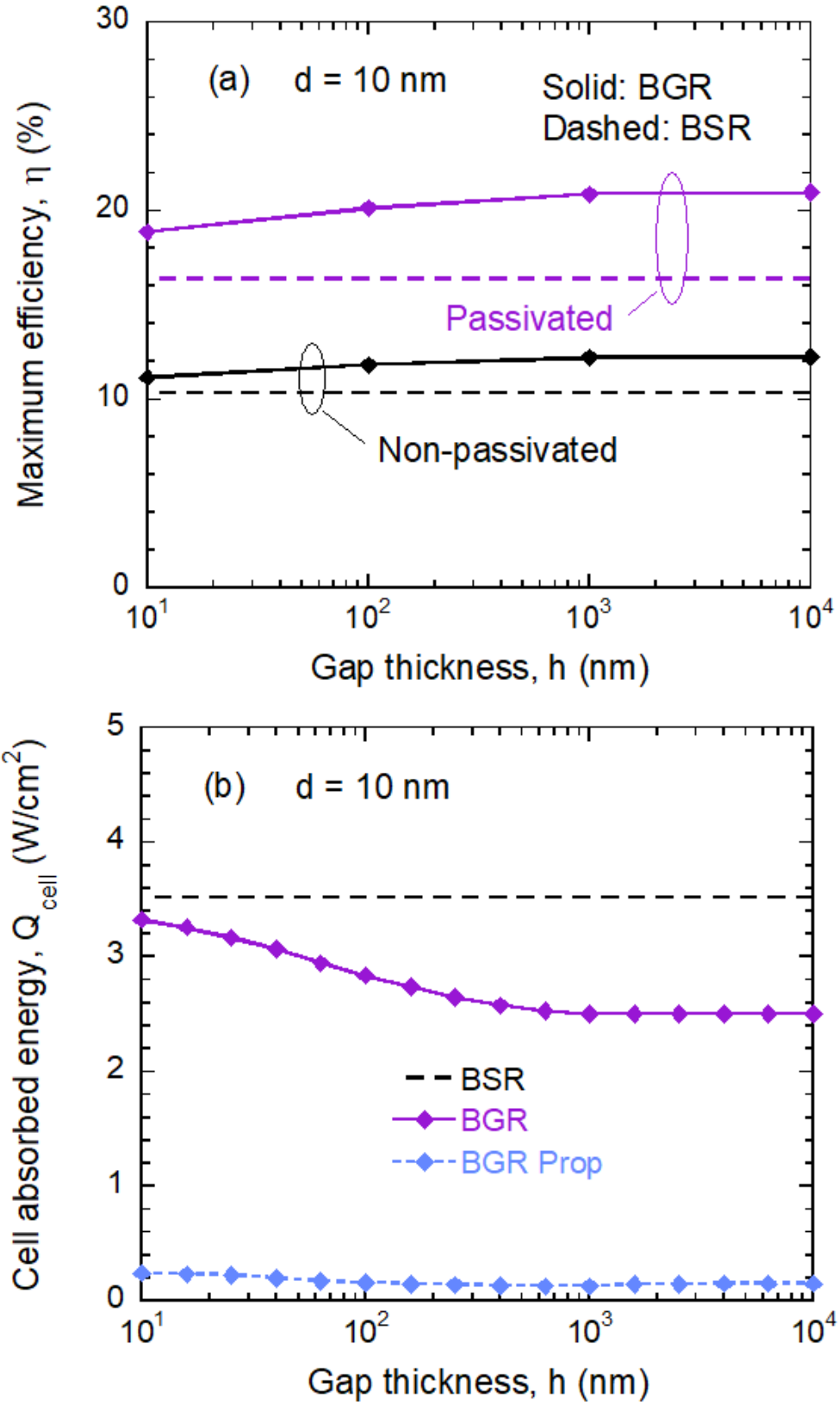

Fig. 5, Feng et al.

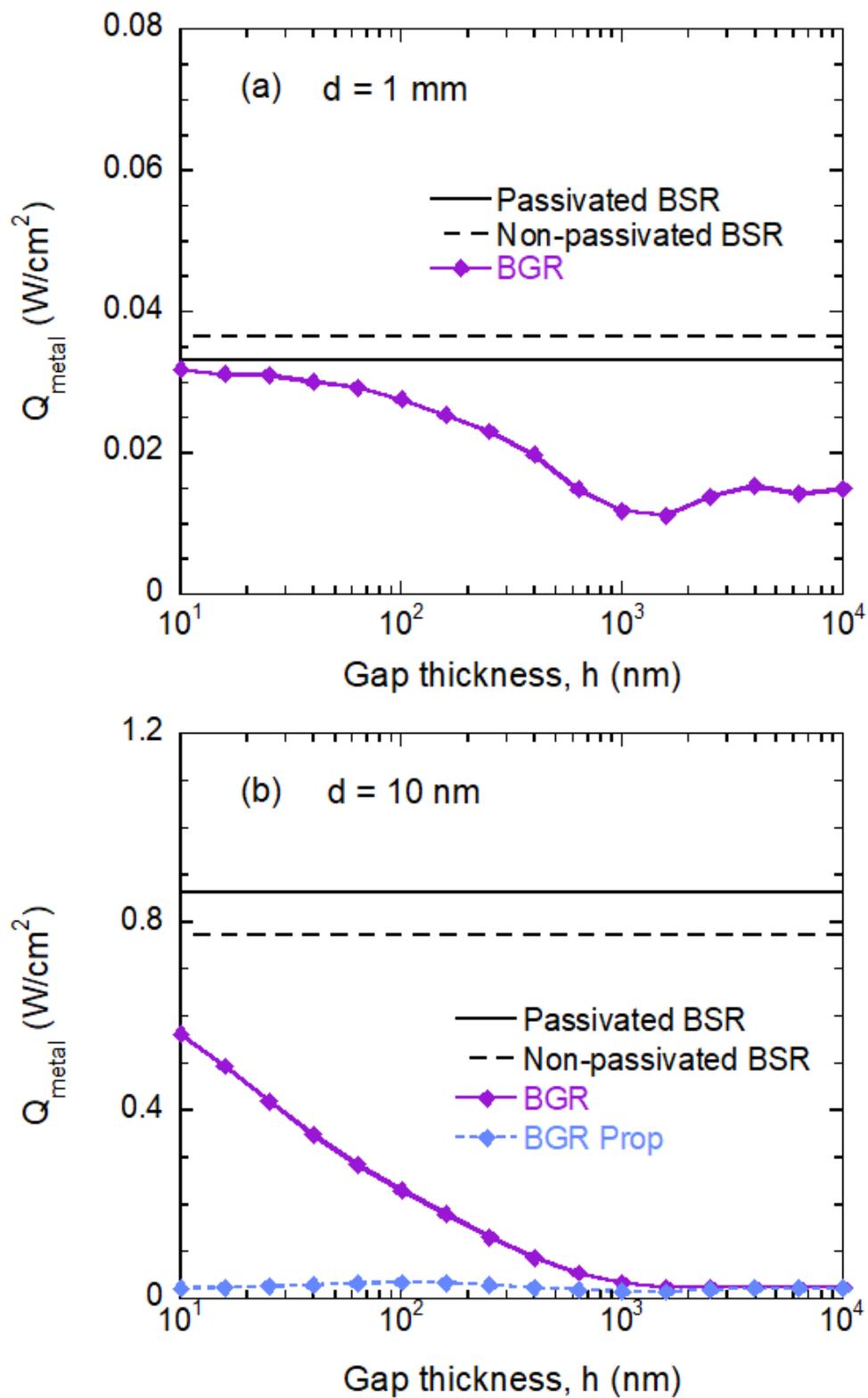

Fig. 6, Feng et al.

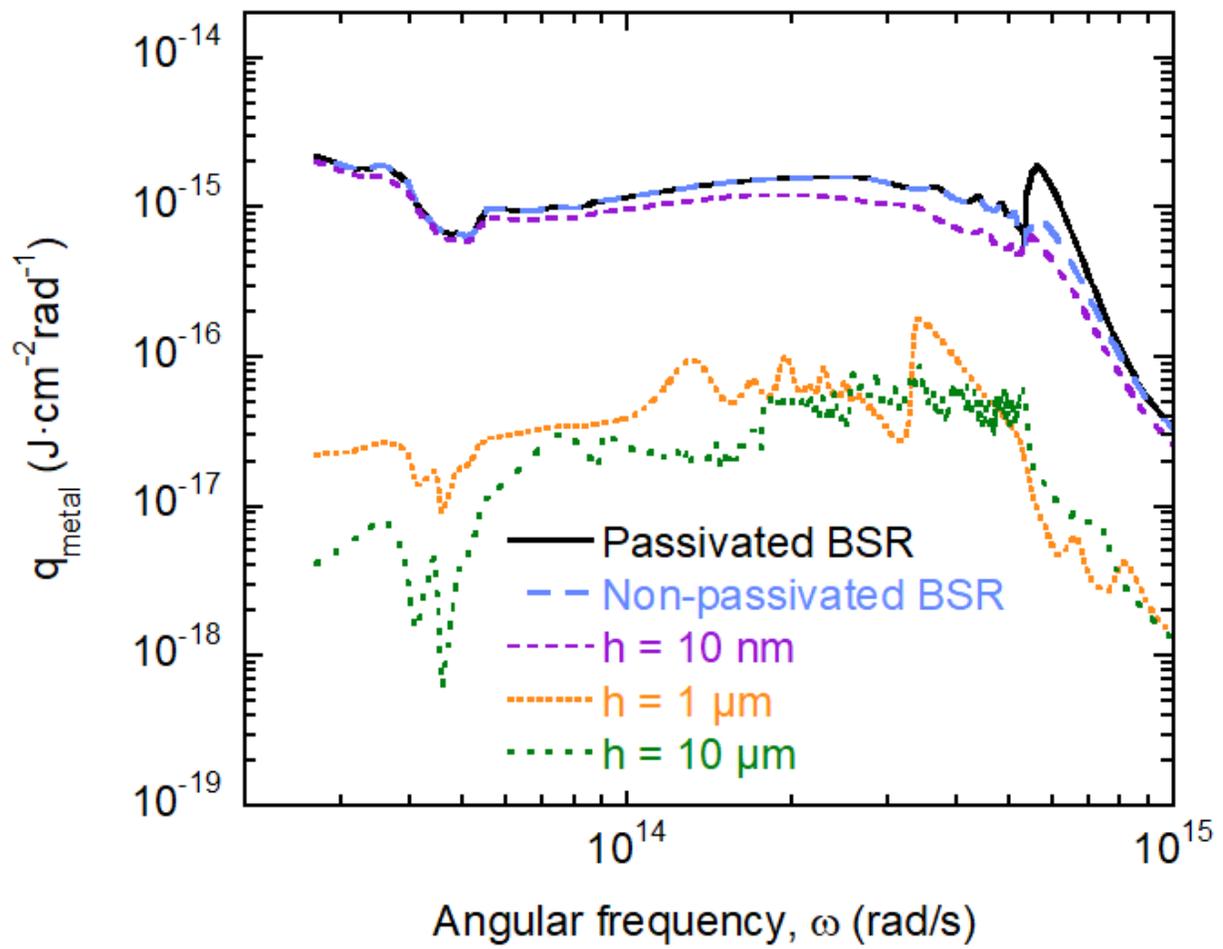

Fig. 7, Feng et al.

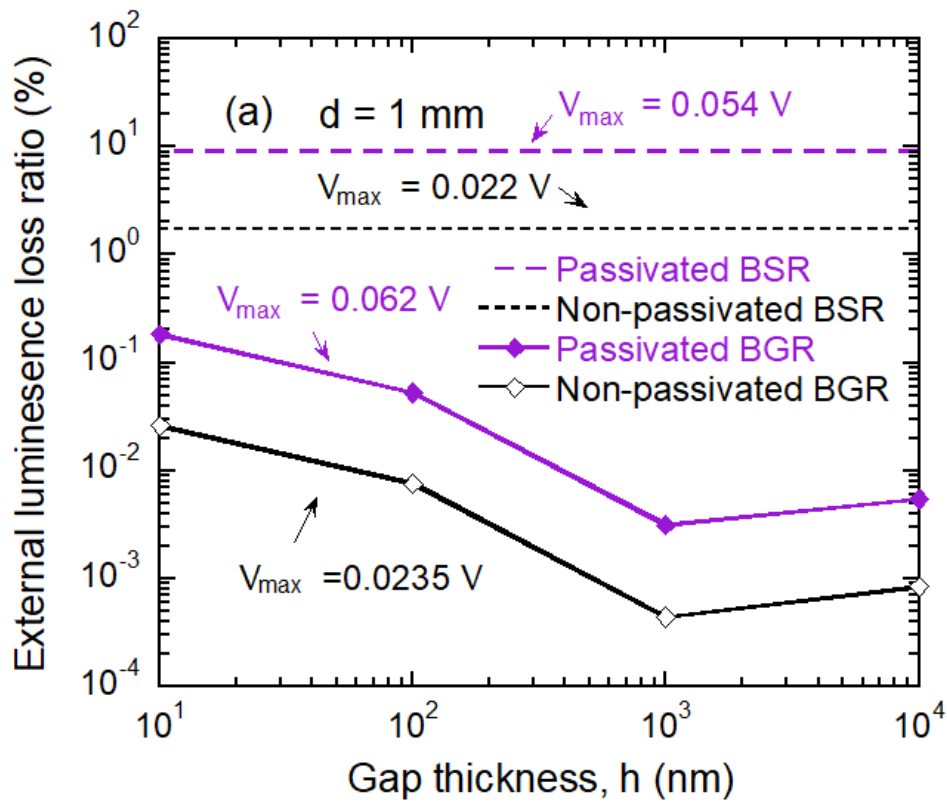

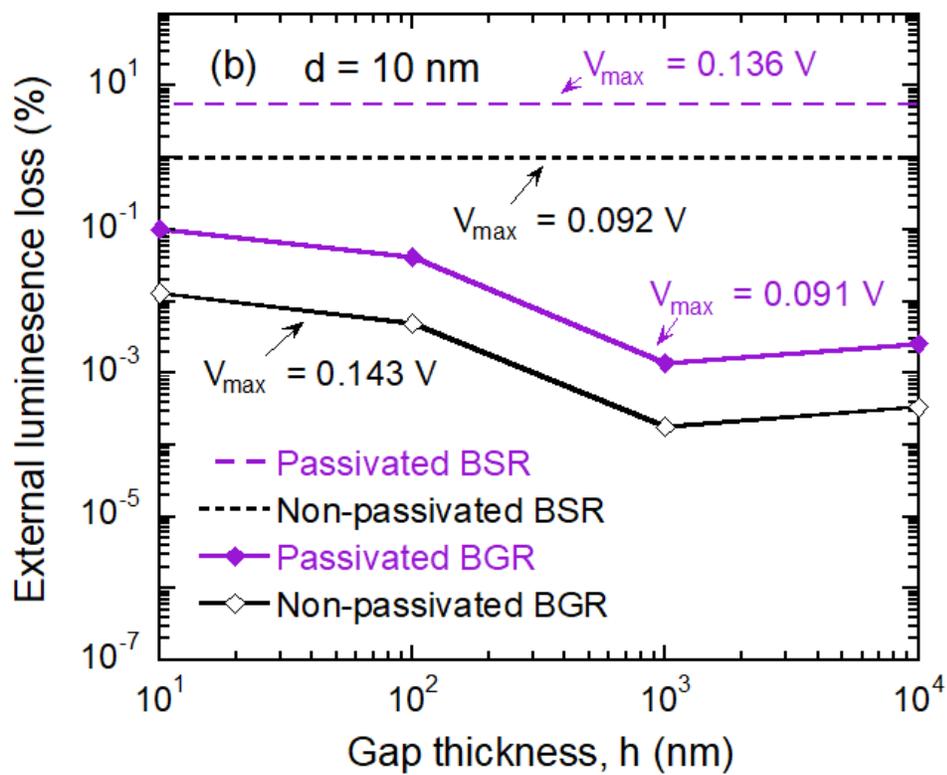

Fig. 8, Feng et al.